\documentclass[sigconf, nonacm]{acmart}

\newcommand\vldbyear{2026}
\newcommand\vldbworkshop{DASHSys: Systems for Data-centric Agents with Human-in-the-loop}
\newcommand\vldbauthors{\authors}
\newcommand\vldbtitle{\shorttitle}
\newcommand\vldbavailabilityurl{}   
\newcommand\vldbpagestyle{empty}

\newcommand{\stitle}[1]{\noindent{\textbf{#1.}}}

\usepackage{xcolor}

\newcommand{\chip}[3]{%
  {\setlength{\fboxsep}{1.2pt}%
  \colorbox{#2}{\textcolor{#3}{\scriptsize\textsf{\textbf{#1}}}}}%
}

\newcommand{\plog}{\chip{P1}{orange!18}{orange!70!black}}
\newcommand{\pflow}{\chip{P2}{blue!14}{blue!65!black}}
\newcommand{\pstoch}{\chip{P3}{purple!14}{purple!65!black}}

\newcommand{\srecon}{\chip{S1}{green!14}{green!50!black}}
\newcommand{\sjoin}{\chip{S2}{red!12}{red!65!black}}

\usepackage{xspace}
\newcommand{\name}{\textsc{AgentTrails}\xspace}

\newcommand\revision[1]{\textcolor{black}{#1}}

\begin{document}
\title{\name: Towards Trust and Reuse for Agentic Tasks }






\author{Eden Wu, Sonia Castelo, Yurong Liu, Cl\'{a}udio T. Silva and Juliana Freire}
\affiliation{%
  \institution{New York University}
}
\email{{eden.wu, s.castelo, yurong.liu,  csilva, juliana.freire}@nyu.edu}

\begin{abstract}
  LLM-powered agents increasingly tackle complex tasks by invoking tools, querying databases, executing code, and manipulating intermediate artifacts. These agents follow trajectories that are typically stored as chronological logs, obscuring the underlying dataflow -- the dependencies between their actions and the artifacts they create and manipulate. This limits developers’ ability to understand the agents' trails, 
  compare executions, debug failures, and re-use the computations.
%
We present \name, a prototype system for agent provenance and sensemaking. \name converts raw trajectories into structured provenance graphs, where tool calls are modeled as computational actions and inputs and outputs as data artifacts. The system supports the comparison of executions by placing multiple provenance graphs on a shared canvas and constructing a joined quotient graph that aligns recurring tools, artifacts, and dependency structures across trajectories. On top of this representation, \name supports pattern extraction, downstream analysis, and skill abstraction.
We demonstrate \name on real-world agent trajectories, showing that it reveals hidden dependencies, aligns divergent executions, and surfaces recurring tool-use patterns beyond chronological logs.

\end{abstract}

\maketitle

\pagestyle{\vldbpagestyle}
\begingroup\small\noindent\raggedright\textbf{VLDB Workshop Reference Format:}\\
\vldbauthors. \vldbtitle. VLDB \vldbyear\ Workshop: \vldbworkshop.\\
\endgroup
\begingroup
\renewcommand\thefootnote{}\footnote{\noindent
This work is licensed under the Creative Commons BY-NC-ND 4.0 International License. Visit \url{https://creativecommons.org/licenses/by-nc-nd/4.0/} to view a copy of this license. For any use beyond those covered by this license, obtain permission by emailing \href{mailto:info@vldb.org}{info@vldb.org}. Copyright is held by the owner/author(s). Publication rights licensed to the VLDB Endowment. \\
\raggedright Proceedings of the VLDB Endowment. ISSN 2150-8097. \\
}\addtocounter{footnote}{-1}\endgroup

\ifdefempty{\vldbavailabilityurl}{}{
\vspace{.3cm}
\begingroup\small\noindent\raggedright\textbf{VLDB Workshop Artifact Availability:}\\
The source code, data, and/or other artifacts have been made available at \url{\vldbavailabilityurl}.
\endgroup
}

\vspace{-.15cm}
\section{Introduction}
\vspace{-.05cm}

LLM-powered agents are increasingly used in domains such as software engineering, scientific discovery, and data analysis, where they execute tasks through sequences of tool calls, external queries, code execution, and intermediate artifact manipulation~\cite{yao2023react,wei2025aiscienceagenticscience}. These executions produce rich trajectories containing messages, tool invocations, responses, generated files, and intermediate results that can be useful for debugging failures, analyzing tool-use behavior, comparing agents and models, and curating data for post-training. Repeated structures across successful executions may further reveal reusable skills or workflows.
However, because the trajectories are stored as an unstructured, sequential log, they obscure the workflow the agent orchestrates, making it difficult to understand the underlying logic of the agent's actions.


\stitle{Understanding Agent Traces: Challenges} Raw agent trajectories are typically recorded as chronological textual logs. While they preserve what happened, they are difficult to use for understanding, comparing, and improving agent behavior:

\plog~\textit{Raw traces are long and heterogeneous.} A single trace may contain many turns, tool schemas, structured arguments, responses, and generated artifacts; comparing traces as text is tedious and offers no compact overview of tool usage or execution patterns.
%
\pflow~\textit{Chronological order hides workflow structure.} Later tool calls may reuse artifacts produced many steps earlier, combine outputs from multiple calls, or branch from intermediate results. Conversely, adjacent calls may be unrelated. Chronological order alone, therefore, does not reveal dependencies, artifact reuse, or workflow structure~\cite{souza2025provagent}.
%
\pstoch~\textit{Agent executions are stochastic.} The same task can produce different tool orders, repeated calls, or divergent branches across runs, agents, or models~\cite{yao2024tau,bjarnason2026randomnessagenticevals,chen2025reasoningmodelsdontsay,yuan2026understanding,Kapoor2024AIAT}. This makes it difficult to compare trajectories directly as sequences.

\stitle{From Traces to Provenance: Challenges}
A natural abstraction for agent behavior is execution provenance. However, constructing provenance representations from raw traces introduces two additional challenges:

\srecon~\textit{Raw traces obscure dependency topology.} Agent trajectories rarely contain explicit dependency edges. Recovering provenance therefore requires identifying which tool inputs depend on which prior outputs from heterogeneous evidence such as artifact identifiers, paths, query strings, and semantic references embedded in responses. Reconstruction must be efficient, auditable, and robust across diverse tools.
%
\sjoin~\textit{Multiple provenance graphs lack common alignment anchors.} Comparing provenance graphs requires more than matching tool names or positions: traces may contain repeated calls, missing steps, or alternative branches, so similar workflow stages appear in different structural contexts. Multi-trace analysis requires abstractions that align activities by workflow role and dependency structure.

\stitle{Our Approach: \name}We present \name, a prototype visual analytics system for tool-calling agent provenance and sensemaking. \name provides three coordinated levels of analysis. First, to address \plog~and provide a holistic entry point, \name offers an UpSet-style overview of tool usage across traces. The view shows which traces invoke which tools, how often tools are called, and how tool usage relates to trace-level metrics such as score, cost, latency, or token count. This enables users to filter, group, and select traces before inspecting detailed provenance. Second, to address \pflow~and \srecon, \name converts a selected raw trajectory into a structured provenance graph. Tool calls are represented as computational actions, while their inputs, outputs, intermediate artifacts, and returned values are represented as data artifacts. The graph is constructed from trace-derived evidence such as arguments, responses, artifact identifiers, paths, filenames, URLs, and reused semantic values. Third, to address \pstoch~and \sjoin, \name supports multi-trace comparison through a joined quotient graph. It abstracts tool calls into activity capsules, clusters similar capsules across traces, and aligns recurring tools, artifacts, and dependency structures while preserving trace-specific branches.

\name includes an LLM-based analysis agent that inspects the reconstructed provenance graphs and joined quotient graphs to summarize observed patterns and explain selected branches. The agent supports lightweight downstream sensemaking, while the core workflow remains grounded in trace-derived structures.

This paper makes three preliminary contributions:
i) A provenance framing for tool-calling agent sensemaking, focused on recovering dependency topology from chronological traces. 
ii) A trace-agnostic approach that extracts and infers dependencies from structural and semantic evidence.
iii) A joined provenance abstraction for aligning multiple traces and exposing shared workflows, divergent branches, and low-support behaviors.
\vspace{-.15cm}
\section{Related Works}
\vspace{-.05cm}
\stitle{Workflows and Provenance} Provenance has been studied in the context of scientific workflow systems
to capture workflow evolution, execution context, and data lineage with the goal of supporting reproducibility and reuse~\cite{freire-cise2008}. Approaches have been proposed 
that use visualization to make sense of provenance, including summaries of workflow collections~\cite{koop2013visual} and to compare machine learning pipelines~\cite{ono2021pipelineprofiler}.
%
%
These systems assume workflows are explicitly defined or recoverable from controlled environments, while tool-calling agent trajectories arrive as chronological logs with implicit dependencies and stochastic executions. \name addresses this gap by focusing on provenance reconstruction and multi-trace analysis of heterogeneous agent trajectories.

\stitle{Agent Trajectory Analysis, Provenance, and Reuse} Recent work has explored \emph{agent trajectories} as objects of analysis. PROV-AGENT models provenance for agentic workflows\revision{\ by instrumenting the running framework, one graph per run}~\cite{souza2025provagent}; CHIEF, TRAIL, and AgenTracer transform flat logs into structured representations for debugging~\cite{wang2026flatlogscausalgraphs,deshpande2025trailtracereasoningagentic,zhang2025agentracerinducingfailurellm}; and AgentLens, Agent Trajectory Explorer, SeaView, and Graphectory support visualization and process-level analysis of agent behavior~\cite{lu2024agentlensvisualanalysisagent,desmond2025agenttrajectoryexplorer,bula2025seaviewsoftwareengineeringagent,liu2025graphectory}.
These efforts largely focus on single-run inspection or domain-specific settings.
%
In contrast, \name \revision{reconstructs provenance post-hoc from raw logs without instrumentation and---uniquely---aligns multiple executions through a joined graph}, supporting downstream pattern extraction and sensemaking.
On the reuse front, ExpeL, Reflexion, and Trace2Skill use prior trajectories for workflow induction, reflection, or skill extraction~\cite{zhao2024expel,shinn2023reflexion,ni2026trace2skill}. \name is complementary: rather than reusing trajectories as flat sequences, it exposes their internal dependency structure, enabling identification of reusable motifs while pruning redundant tool calls.
%

\vspace{-.15cm}
\section{System Overview}
\vspace{-.05cm}


\begin{figure*}[t]
    \centering
    \includegraphics[width=0.88\textwidth]{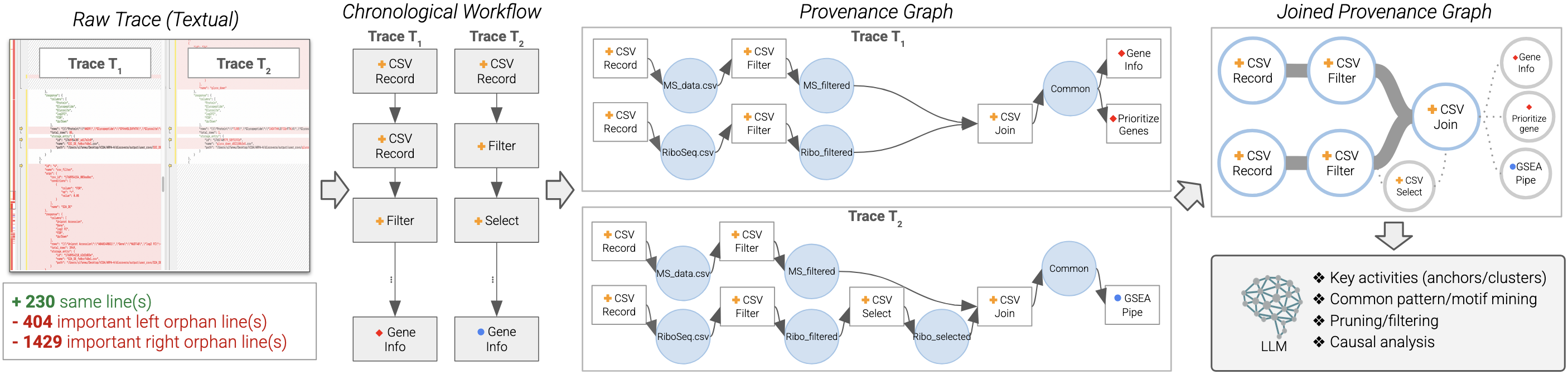}
    \vspace{-.1cm}
    \caption{\name abstracts raw traces into chronological workflows that expose the sequence of actions, then reconstructs a provenance graph that represents actions as activities and artifacts as entities. For multi-trace analysis, \name aligns similar activities across provenance graphs into a single joined provenance graph, enabling downstream tasks.}
    \vspace{-.15cm}
    \label{fig:trace-to-provenance}
\end{figure*}



\name targets raw agent traces stored as long, heterogeneous event logs. To address \plog, it normalizes each trace into a sequence of actions while preserving evidence fields such as arguments, responses, artifact identifiers, paths, URLs, returned objects, and reused semantic values. This design does not assume a fixed agent architecture or tool schema, allowing \name to support traces from heterogeneous frameworks and domains.

\stitle{Trace Overview with Tool Coverage} The first stage provides a global, tool-agnostic overview of selected traces. 
For a trace \(T_r=\langle c_{r,1},\ldots,c_{r,n_r}\rangle\)\revision{, where \(c_{r,i}\) denotes the \(i\)-th tool call of trace \(r\) together with its arguments and response,}  and tool set \(\mathcal{U}\), \name builds a coverage matrix where each entry \(M_{r,u}=|\{i:\mathrm{tool}(c_{r,i})=u\}|\) counts how many times tool \(u\) is invoked in trace \(r\).
Rows encode traces, columns encode tools, and cell values encode repeated calls. The matrix is shown as an UpSet-style overview (Fig.~\ref{fig:discovera-scenario}A), with a top bar chart for total tool frequency or Shapley-style tool impact, and a side rail for trace-level metrics such as score, cost, latency, or token count. This view does not infer dependencies; it supports holistic filtering, grouping, and trace selection before graph-level analysis.

\stitle{Provenance Graph Construction}
To address \pflow~and~\srecon, \name converts a selected chronological trace into a provenance graph that makes producer-consumer dependencies explicit (Fig.~\ref{fig:trace-to-provenance}). For a trace \(T\), the graph is
\revision{\(G_T=(A_T \cup E_T,\; R_T)\)},
where \(A_T\) contains tool-call activities and \(E_T\) contains recovered entities such as inputs, outputs, artifacts, and returned values. \revision{The edge set \(R_T \subseteq (A_T \times E_T) \cup (E_T \times A_T)\) keeps the graph bipartite: \emph{generatedBy} edges are always emitted from each activity to its outputs, while \emph{usedBy} and weaker \emph{informedBy} edges---the reconstruction targets---link entities to later activities that consume them.}
The key difficulty is that raw traces rarely provide dependency edges directly. \name therefore treats graph construction as evidence-based reconstruction. It first creates a deterministic skeleton from exact evidence: output entities are extracted from response metadata such as identifiers, paths, filenames, URLs, names, or returned objects. A dependency is added when a later call explicitly references an earlier entity: \vspace{-.2cm}
\[
e_j \rightarrow a_i
\quad \text{if} \quad
K(e_j)\cap \mathrm{refs}(c_i.\mathrm{args})\neq \emptyset,\; j<i,
\]
where \(K(e_j)\) denotes the recovered keys for entity \(e_j\)\revision{---identifying strings from its producing response, e.g., a storage id, filename, or URL}. This step captures high-precision artifact reuse while enforcing temporal validity\revision{, keeping the graph acyclic: agent loops surface as repeated activities, which the joined graph aggregates with per-trace multiplicity}.

\name also extracts weaker semantic evidence from shared values, query terms, table columns, domain objects, or tokens that appear in earlier responses and later arguments. These matches are stored as dependency candidates rather than asserted as ground truth. A constrained LLM refinement step receives the draft graph and candidate relations, then returns graph patch operations. Only patches that preserve valid node references, temporal order, and the activity/entity schema are accepted. This design keeps the provenance graph auditable: exact edges, semantic candidates, and LLM-refined edits remain distinguishable.

Two questions remain for scaling this methodology: provenance quality needs gold dependency benchmarks, and asking LLMs to infer dependencies directly from full traces does not scale. \name therefore retrieves likely candidates first---via indexed artifact keys, value sketches, schema-aware blocking, and semantic matching---and uses an LLM only to refine or explain ambiguous ones. \revision{As a first step, we hand-annotated 10 traces with 234 gold dependency edges; preliminary results are promising and a full staged evaluation is underway.}

\begin{figure}[t]
    \centering
    \includegraphics[width=0.92\columnwidth]{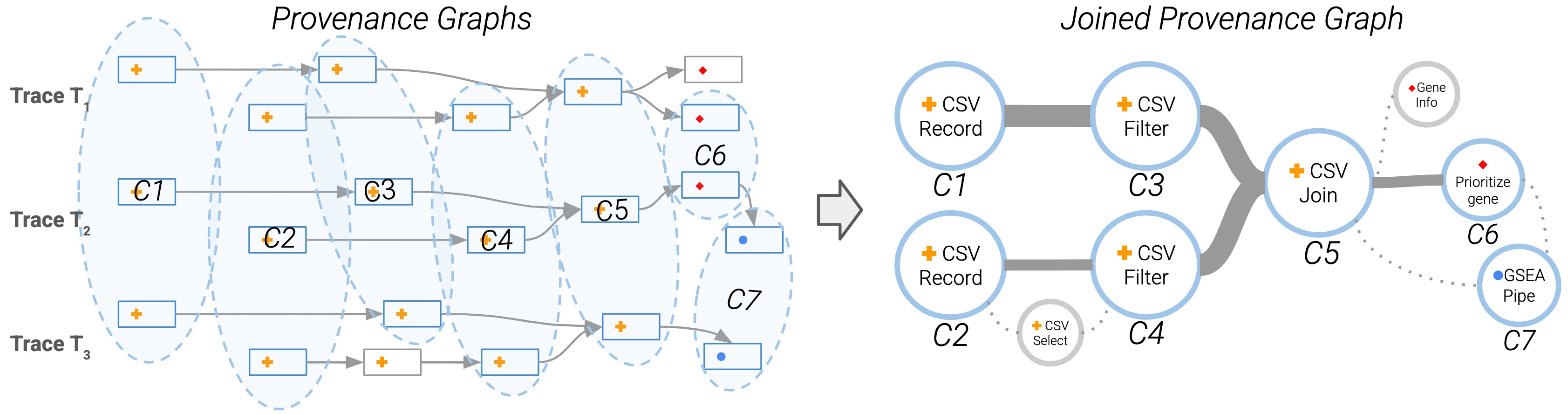}
        \vspace{-.2cm}

    \caption{\name aligns similar activities across multiple provenance graphs into clusters (left), then merges each cluster into a joined anchor (right).}
    \label{fig:prov-to-quotient}
    \vspace{-.5cm}
\end{figure}

\stitle{Multi-Trace Provenance Graph}
To address \pstoch~and \sjoin, \name constructs a joined provenance graph over selected traces. This stage targets the alignment problem: different traces may solve the same task with different tool orders, repeated calls, missing steps, or alternative branches. Thus, comparing traces by raw sequence or exact tool name is insufficient.

As shown in Fig.~\ref{fig:prov-to-quotient}, \name first abstracts each activity into an activity capsule, which compactly describes the activity's tool identity, input/output structure, local graph context, and upstream root lineage. Capsules are then clustered across traces using a weighted similarity over these features:     \vspace{-.15cm}
\[
S(\kappa_i,\kappa_j)=\sum_m w_m\,s_m(\kappa_i,\kappa_j),
\]    
where \revision{\(\kappa_i,\kappa_j\) are capsules and each \(s_m\in[0,1]\) compares one feature---tool tokens, input/output structure, graph context, root lineage, or evidence keywords---via Jaccard or cosine similarity, with weights \(w_m\) summing to one}. Each cluster becomes a joined activity node. Original provenance edges are then remapped based on the cluster assignments: if two original edges connect activities assigned to the same source and target clusters, they are aggregated into one joined edge.

Joined nodes and edges store supporting traces, support counts, per-trace multiplicity, representative members, and optional score summaries. High-support structures reveal recurring workflow motifs, while low-support structures expose trace-specific branches, redundant operations, or anomalous paths. The joined graph therefore summarizes a set of runs at the provenance level rather than as an aggregate tool sequence.

This abstraction raises a follow-up question: joined provenance construction can be designed in many ways. Prior work on workflow analogies suggests that graphs may be aligned by operators, data dependencies, execution roles, or higher-level workflow intent~\cite{scheidegger2007analogy}. \name takes a first step by clustering activity capsules into joined anchors, but future work should explore alternative abstractions and scalable candidate alignment through blocking, retrieval, and role-aware graph features.

\stitle{Interactive Visualization, Filtering, and Copilot}
\name provides coordinated views for overview, inspection, and comparison. The UpSet view supports trace selection, metadata grouping, score comparison, and tool-usage analysis. The single-trace view exposes recovered activities, entities, arguments, responses, and edge evidence. The joined view compares multiple traces, using node size for activity support, edge width for dependency support, and color to preserve trace membership. Users can fade or prune nodes and edges by support or score, enabling them to identify dominant workflows, low-support anomalies, and differences between high- and low-scoring traces.

In addition to these views, \name includes a provenance copilot for lightweight interactive sensemaking. Given the current selection, the copilot can inspect graph structures, retrieve tool inputs and outputs, expand joined nodes to their trace members, and summarize visible patterns. The copilot does not define provenance; it helps users interpret trace-derived, auditable graphs.



\vspace{-.15cm}
\section{Usage Scenarios}
\vspace{-.05cm}

\stitle{Inspecting a SciAgentGym Physics Trace}
%
We first demonstrate \name on a SciAgentGym task. SciAgentGym evaluates multi-step scientific tool use by LLM agents and provides more than 1,780 tools across Physics, Chemistry, Materials Science, Life Science, and Astronomy~\cite{shen2026sciagentgym}. We analyze task \texttt{27} from Physics, which asks for the hyperfine transition frequency of a hypothetical ground-state hydrogen atom with electron spin \(\frac{3}{2}\). The raw trace contains structured tool calls, numerical outputs, and generated visualizations, making the scientific workflow difficult to verify from text or chronological order alone.

\begin{figure}[t]
    \centering
    \includegraphics[width=\columnwidth]{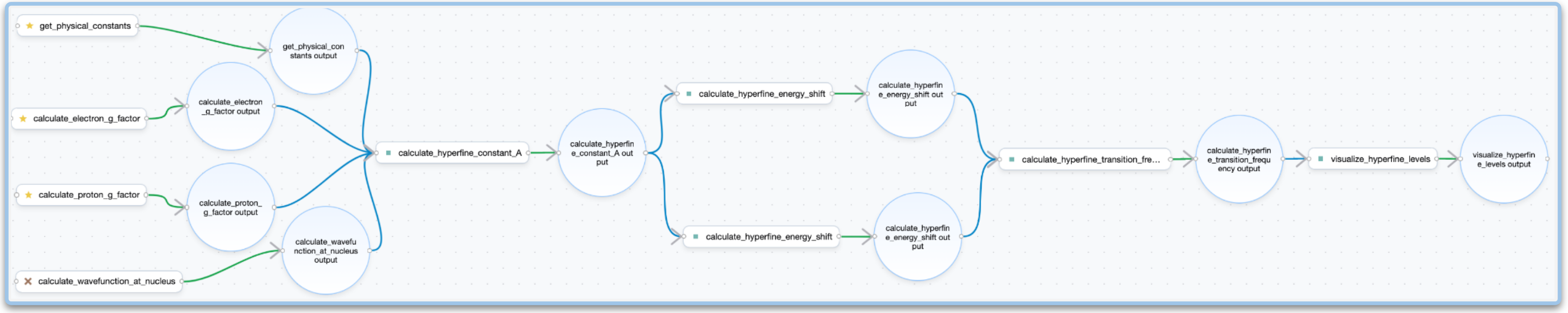}
    \vspace{-.35cm}
    \caption{SciAgentGym usage scenario. \name exposes the dependency structure hidden in a raw physics trace.}
    \label{fig:sciagentgym-scenario}
    \vspace{-.05cm}
\end{figure}

After loading the trace into \name, the reconstructed provenance graph exposes a clear fan-in/fan-out computation pattern (Fig.~\ref{fig:sciagentgym-scenario}). Independent quantities-physical constants, electron and proton \(g\)-factors, and the wavefunction value at the nucleus-converge into the calculation of the hyperfine constant \(A\). This intermediate result then branches into two energy-shift calculations for the \(F=1\) and \(F=2\) states. These shifts, together with the angular-momentum states, support the final transition-frequency calculation and the visualization outputs. This dependency structure is not apparent from the raw sequential trace, where related calls may be separated and where the scientific role of each intermediate output is buried in tool responses.

Comparing the reconstructed graph with the benchmark's expected tool use shows that the trace covers the required scientific steps. More importantly, the provenance graph reveals how those steps are connected: which upstream quantities feed the hyperfine constant, how \(A\) is reused in multiple downstream computations, and how the final frequency is grounded in earlier tool results. This scenario demonstrates how \name supports single-trace validation and scientific workflow inspection, rather than benchmarking the underlying model.

\begin{figure}[t]
    \centering
    \includegraphics[width=\columnwidth]{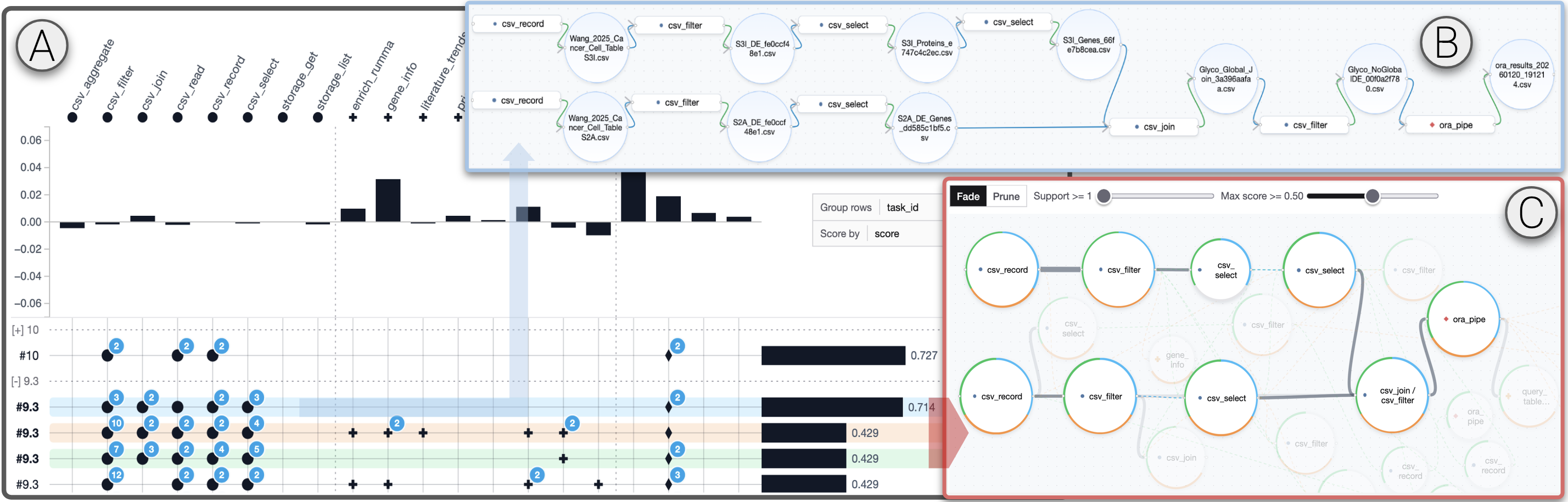}
    \vspace{-.35cm}
    \caption{Discovera usage scenario. \name helps users move from trace-level overview (A), to single-trace provenance inspection (B), to joined multi-trace comparison (C).}
    \Description{Interface screenshot showing a trace overview, a single-trace provenance graph, and a joined provenance graph for comparing Discovera runs.}
    \label{fig:discovera-scenario}
    \vspace{-.5cm}
\end{figure}

\stitle{Discovera Gene-Set Exploration}
%
We also demonstrate \name on Discovera traces for Signature-to-Mechanisms analysis. Discovera is a workflow-aligned scientific agent in which the LLM orchestrates deterministic tools, stores intermediate artifacts, and grounds reasoning in tool outputs
~\cite{pintoveizaga2026discovera}. We analyze task \texttt{s2m\_task\_016}. In the UpSet overview, we group runs by task ID and rank traces by score (Fig.~\ref{fig:discovera-scenario}A). The top trace uses fewer tool calls than several alternatives, motivating provenance-level inspection.

Opening the top trace reveals a compact workflow: two Wang 2025 Cancer Cell datasets are ingested with \texttt{csv\_record}, filtered by \texttt{FDR < 0.05}, projected through \texttt{csv\_select}, joined with \texttt{csv\_join}, refined to remove global differentially expressed genes, and analyzed with \texttt{ora\_pipe}. The provenance graph exposes this as a coherent path from data ingestion to enrichment (Fig.~\ref{fig:discovera-scenario}B).

We then select the next two ranked traces and construct a joined provenance graph. Filtering the joined view by score highlights the high-scoring execution as a direct, well-supported dataflow, while lower-scoring traces contain branches and detours away from the core data-processing path (Fig.~\ref{fig:discovera-scenario}C). Through node support, edge support, and trace membership encodings, \name helps distinguish shared workflow structure from trace-specific deviations\revision{: the lower-scoring runs introduce extra calls off the dominant provenance path---concrete candidates for pruning or prompt fixes}.


\section{Conclusion}

AgentTrails is preliminary work toward a principled infrastructure for agent provenance and sensemaking. We presented a provenance framing for tool-calling agent traces, evidence-based dependency-graph reconstruction from chronological logs, and a joined quotient graph for aligning and comparing executions. Usage scenarios on SciAgentGym and Discovera traces demonstrate that the system surfaces dependency structure and workflow patterns that are not recoverable from raw sequential logs. Several important questions remain open: provenance quality lacks ground-truth benchmarks; the capsule similarity function and its weights require principled tuning and evaluation; and scalability to very long traces or large trace collections has not been assessed. We view \name as a foundation for future work on agent debugging, workflow reuse, and skill extraction grounded in auditable, trace-derived provenance.

\vspace{-.05cm}
\begin{acks}
\revision{This work was supported in part by DARPA ASKEM (HR0011262087), ARPA-H BDF, and NSF (OAC-2411221). The views, opinions, and findings expressed are those of the authors and should not be interpreted as representing the views or policies of these agencies.}
\end{acks}


\bibliographystyle{ACM-Reference-Format}
\bibliography{main}

@article{freire-cise2008,
author = {Juliana Freire and David Koop and Emanuele Santos and Cl\'audio T. Silva},
title = {{Provenance for Computational Tasks: A Survey}},
journal ={Computing in Science and Engineering},
volume = {10},
number = {3},
issn = {1521-9615},
year = {2008},
pages = {11-21},
OPTdoi = {http://doi.ieeecomputersociety.org/10.1109/MCSE.2008.79},
publisher = {IEEE Computer Society},
OPTaddress = {Los Alamitos, CA, USA},
}

@misc{wei2025aiscienceagenticscience,
  author = {Jiaqi Wei and Yuejin Yang and Xiang Zhang and others},
  title = {From AI for Science to Agentic Science: A Survey on Autonomous Scientific Discovery},
  year = {2025},
  eprint = {2508.14111},
  archivePrefix = {arXiv},
}

@inproceedings{
yao2024tau,
title     = {$\tau$-bench: A Benchmark for Tool-Agent-User Interaction in Real-World Domains},
author={Shunyu Yao and Noah Shinn and Pedram Razavi and others},
booktitle={ICLR},
year={2025}
}

@inproceedings{bjarnason2026randomnessagenticevals,
  title     = {On Randomness in Agentic Evals},
  author    = {Bjarni Haukur Bjarnason and Andr{\'e} Silva and Martin Monperrus},
  booktitle = {ICLR 2026 Workshop on Agents in the Wild},
  year      = {2026}
}

@misc{chen2025reasoningmodelsdontsay,
  author = {Yanda Chen and Joe Benton and Ansh Radhakrishnan and others},
  title = {Reasoning Models Don't Always Say What They Think},
  year = {2025},
  eprint = {2505.05410},
  archivePrefix = {arXiv},
}

@inproceedings{yuan2026understanding,
  author = {Jiayi Yuan and Hao Li and Xinheng Ding and others},
  title = {Understanding and Mitigating Numerical Sources of Nondeterminism in {LLM} Inference},
  booktitle = {NeurIPS},
  year = {2025},
}

@article{Kapoor2024AIAT,
  author = {Sayash Kapoor and Benedikt Stroebl and Zachary S. Siegel and others},
  title = {AI Agents That Matter},
  journal = {Trans. Mach. Learn. Res.},
  year = {2024},
  volume = {2025},
}

@inproceedings{yao2023react,
  title     = {ReAct: Synergizing Reasoning and Acting in Language Models},
  author    = {Shunyu Yao and Jeffrey Zhao and Dian Yu and Nan Du and Izhak Shafran and others},
  booktitle = {ICLR},
  year      = {2023}
}

@inproceedings{koop2013visual,
  author = {Koop, David and Freire, Juliana and Silva, Cláudio T.},
  title = {Visual summaries for graph collections},
  booktitle = {IEEE PacificVis},
  year = {2013},
  doi = {10.1109/PacificVis.2013.6596128},
}

@article{ono2021pipelineprofiler,
  author = {Jorge Piazentin Ono and Sonia Castelo and Roque L{\'o}pez and others},
  title = {PipelineProfiler: A Visual Analytics Tool for the Exploration of AutoML Pipelines},
  journal = {IEEE Trans. Vis. Comput. Graph.},
  year = {2020},
  volume = {27},
  url = {https://api.semanticscholar.org/CorpusID:218470098},
}

@INPROCEEDINGS{souza2025provagent,
  author={Souza, Renan and Gueroudji, Amal and DeWitt, Stephen and others},
  booktitle={2025 IEEE International Conference on eScience (eScience)}, 
  title={PROV-AGENT: Unified Provenance for Tracking AI Agent Interactions in Agentic Workflows}, 
  year={2025},
  volume={},
  number={},
  pages={467-473}}

@misc{wang2026flatlogscausalgraphs,
  author = {Yawen Wang and Wenjie Wu and Junjie Wang and others},
  title = {From Flat Logs to Causal Graphs: Hierarchical Failure Attribution for LLM-based Multi-Agent Systems},
  year = {2026},
  eprint = {2602.23701},
  archivePrefix = {arXiv},
}

@misc{deshpande2025trailtracereasoningagentic,
  author = {Darshan Deshpande and Varun Gangal and Hersh Mehta and others},
  title = {TRAIL: Trace Reasoning and Agentic Issue Localization},
  year = {2025},
  eprint = {2505.08638},
  archivePrefix = {arXiv},
}

@inproceedings{
zhang2025agentracerinducingfailurellm,
title={AgenTracer: Who Is Inducing Failure in the {LLM} Agentic Systems?},
author = {Guibin Zhang and Junhao Wang and Junjie Chen and Wangchunshu Zhou and others},

booktitle={ICLR},
year={2026}
}

@article{lu2024agentlensvisualanalysisagent,
  author={Jiaying Lu and Bo Pan and others},
  title={AgentLens: Visual Analysis for Agent Behaviors in LLM-Based Autonomous Systems},
  year={2025},
  cdate={1754006400000},
  journal={IEEE Trans. Vis. Comput. Graph.},
  volume={31},
  number={8}
}

@article{desmond2025agenttrajectoryexplorer,
  author = {Desmond, Michael and Lee, Ja Young and Ibrahim, Ibrahim and others},
  title = {Agent Trajectory Explorer: Visualizing and Providing Feedback on Agent Trajectories},
  journal = {Proc. AAAI Conf. Artif. Intell.},
  year = {2025},
  volume = {39},
  number = {28},
  doi = {10.1609/aaai.v39i28.35350},
}

@misc{bula2025seaviewsoftwareengineeringagent,
  author = {Timothy Bula and Saurabh Pujar and Luca Buratti and others},
  title = {SeaView: Software Engineering Agent Visual Interface for Enhanced Workflow},
  year = {2025},
  eprint = {2504.08696},
  archivePrefix = {arXiv},
}

@article{liu2025graphectory,
  author = {Liu, Shuyang and Chen, Yang and Krishna, Rahul and others},
  title = {Process-Centric Analysis of Agentic Software Systems},
  journal = {Proc. ACM Program. Lang.},
  year = {2026},
  volume = {10},
  number = {OOPSLA1},
  doi = {10.1145/3798271},
}

@inproceedings{zhao2024expel,
  author = {Zhao, Andrew and Huang, Daniel and Xu, Quentin and others},
  title = {ExpeL: LLM agents are experiential learners},
  booktitle = {AAAI},
  year = {2024},
  doi = {10.1609/aaai.v38i17.29936},
}

@inproceedings{shinn2023reflexion,
  author = {Shinn, Noah and Cassano, Federico and Gopinath, Ashwin and others},
  title = {Reflexion: Language Agents with Verbal Reinforcement Learning},
  booktitle = {NeurIPS},
  year = {2023},
  eprint = {2303.11366},
  archivePrefix = {arXiv},
}

@misc{ni2026trace2skill,
  author = {Jingwei Ni and Yihao Liu and Xinpeng Liu and others},
  title = {Trace2Skill: Distill Trajectory-Local Lessons into Transferable Agent Skills},
  year = {2026},
  eprint = {2603.25158},
  archivePrefix = {arXiv},
}

@inproceedings{
shen2026sciagentgym,
title={SciAgentGym: Benchmarking Multi-Step Scientific Tool-Use in {LLM} Agents},
author={Yujiong Shen and Yajie Yang and Zhiheng Xi and Binze Hu and others},
booktitle={ICML},
year={2026}
}

@article{scheidegger2007analogy,
  author = {Scheidegger, Carlos and Vo, Huy and Koop, David and others},
  title = {Querying and Creating Visualizations by Analogy},
  journal = {IEEE Trans. Vis. Comput. Graph.},
  year = {2007},
  volume = {13},
  number = {6},
  doi = {10.1109/TVCG.2007.70584},
}

@misc{pintoveizaga2026discovera,
  author = {Pinto Veizaga, Daniela and Santos, A{\'e}cio and Wu, Eden and others},
  title = {Discovera: A Workflow-Aligned AI Agent for Signature-to-Mechanisms Analysis},
  year = {2026},
  note = {NE Agents Day 2026 Workshop Submission, Submission 18},
}

\end{document}